# 160 ps Yb:YAG/Cr:YAG microchip laser

Xiaoyang Guo[1,2*], Koichi Hamamoto[1,3], Shigeki Tokita[1], Junji Kawanaka[1]

[1]Institute of Laser Engineering, Osaka University, 2-6 Yamadaoka, Suita, Osaka 565-0871, Japan

[2]Department of Electronic Science and Engineering, Kyoto University, Kyoto-Daigaku-Katsura, Nishikyo-ku, Kyoto 615-8510, Japan

[3]Mitsubishi Heavy Industries, Ltd., 16-5 Konan 2-chome, Minato-ku, Tokyo, Japan

*Corresponding author: guoxiaoyang@ile.osaka-u.ac.jp

**Abstract:** By cryogenically cooling the Yb:YAG/Cr:YAG medium, one can break through the damage limit of Yb:YAG/Cr:YAG passively Q-switched microchip lasers at room temperature and thus achieve a shorter minimum pulse duration. In the proof of principle experiment we carried out, a 160.6 ps pulse duration was obtained. To the best of our knowledge, this is the first realization of sub-200 ps pulse operation for an Yb:YAG/Cr:YAG microchip laser.

Picosecond lasers are desirable for applications such as industrial micro-material fabrication processes. Currently, the typical method to generate such picosecond pulses is using a mode locked oscillator. However, mode locked oscillators are complex, which makes them expensive to purchase and maintain. In addition, the output pulse energy is low, which limits their application. Usually, passively Q-switched (PQS) lasers are used for generating nanosecond pulses. However, a PQS laser's pulse duration is proportional to the cavity length, so it is possible to use a very short cavity length to achieve picosecond pulse operation. Therefore, PQS microchip lasers have been proposed [1]. PQS microchip lasers are monolithic solid-state lasers in which the gain medium and saturable absorber are in direct contact or bonded, and the cavity mirrors are in direct contact with, or deposited on, the laser medium. The microchip laser cavity length is in the millimeter or even sub-millimeter range, and picosecond pulse generation is possible. Compared with the mode locked oscillator, the PQS microchip laser is simple, cost effective, near alignment free, allows energy scaling and needs less maintenance.

For just over 2 decades, Nd:YVO$_4$/semiconductor saturable absorber mirror (SESAM) and Nd:YAG/Cr:YAG based microchip lasers have been widely investigated. Nd:YVO$_4$/SESAM microchip lasers with a sub-millimeter cavity length can generate sub 100 ps laser pulses [2-5], and the recent record is 16 ps [5]. However, the output energy is typically below 1 μJ, which limits their application. Nd:YAG/Cr:YAG microchip lasers can deliver over 1 μJ, or even 1 mJ, pulse energy [1, 6-9]. However, the absorption cross section of Nd:YAG is much lower than that of Nd:YVO$_4$. If the Nd:YAG thickness is too thin, the transmitted pump will bleach the Cr:YAG absorber and broaden the pulse duration [6]. In addition, unlike SESAM, which acts as a mirror and doesn't contribute the length of cavity, the thickness of Cr:YAG is not to be neglected. All of these limit the Nd:YAG/Cr:YAG microchip laser minimum cavity length. Therefore, the shortest pulse duration obtained directly from a Nd:YAG/Cr:YAG microchip laser is limited to 148 ps [6] and typical pulse durations are over 200 ps [1, 7-9].

Yb:YAG is an attractive candidate for microchip lasers. Compared with Nd:YAG, Yb:YAG has a much higher doping concentration. One can apply a heavy doped Yb:YAG to avoid the pump induced bleaching effect. Thus Yb:YAG/Cr:YAG microchip laser has the potential to get a shorter pulse duration than Nd:YAG/Cr:YAG microchip laser. In addition, Yb:YAG/Cr:YAG microchip laser also can support high energy similar with Nd:YAG/Cr:YAG microchip laser's. Furthermore, compared with Nd:YVO$_4$ and Nd:YAG, Yb:YAG also has the advantages of a much longer storage lifetime and a much broader absorption bandwidth, which reduces the requirements for the diode pump (peak power, bandwidth, etc.). However, the saturation fluence of Yb:YAG is as high as ~9 J/cm$^2$ at room temperature. The generated laser fluence is positively related to the saturation fluence of the gain medium. So with picosecond pulse duration, Yb:YAG microchip laser faces an inevitable damage problem. For that reason, at room temperature, the shortest Yb:YAG/Cr:YAG microchip laser duration is limited to 237 ps [10], to the best of our knowledge.

Cryogenically cooled Yb:YAG has been widely applied in both continuous wave (CW) and pulsed laser systems [11–15]. Recently, some efforts have been made to produce liquid nitrogen (LN) cryogenically cooled Yb:YAG/Cr:YAG PQS microchip lasers [16-18]. However, in the previous cryogenically cooled Yb:YAG lasers, the attention was paid on reducing the threshold pump power and improving the output average power, energy, efficiency, and repetition rate. In this letter, we first propose that by cryogenically cooling the Yb:YAG/Cr:YAG medium, one can break through the damage limit of Yb:YAG/Cr:YAG microchip laser at room temperature and thus achieve a shorter minimum pulse duration. It is based on the fact that at LN temperature (77 K) the Yb:YAG saturation fluence is approximately 5 times lower than at room temperature. Since the material supported damage fluence is nearly proportional to the square root of pulse duration, if we assume 237 ps is the damage limited pulse duration for an Yb:YAG/Cr:YAG microchip laser at room temperature, the damage limited pulse duration will be reduced to 9.5 ps at LN temperature. The proof of principle experiment was carried out. We used a 1.1 mm Yb:YAG/Cr:YAG microchip crystal cryogenically cooled by LN and 160.6 ps (full width at half maximum, FWHM) was obtained. To the best of our knowledge, this is the shortest pulse duration for an Yb:YAG/Cr:YAG microchip laser. By further reducing the Yb:YAG/Cr:YAG thickness and increasing the doping concentration, it is possible to obtain sub-100 ps pulses.

Figure 1 shows a schematic diagram of the microchip laser. An Yb:YAG/Cr:YAG composite crystal fabricated with thermal bonding technology (Cryslaser Inc.) was used in the experiment. The thicknesses of the Yb:YAG and Cr:YAG layers were 0.7 mm and 0.4 mm, respectively. Thus, the total thickness of this microchip crystal was 1.1 mm. The aperture of this microchip crystal was a square with 5 mm sides. Yb:YAG and Cr:YAG were [111] and [110] cut, respectively. The doping concentration of Yb ions was 20 at. % and the initial transmission of the Cr:YAG was 75%. The Yb:YAG surface was coated with a film that is highly reflective (HR, R > 99.8%) at 1030 nm and a film that is anti-reflective (AR, R < 5%) at 940 nm, which worked as an input cavity mirror. The Cr:YAG surface was coated with a partially reflective (PR) film with R = 40% at 1030 nm, which worked as an output coupler. The surface flatness of the coated cavity surfaces was $\lambda/8$ ($\lambda$=633 nm). The microchip crystal was mounted on thermally conductive copper plates. Thin indium foils were used between the crystal and copper contact surfaces to obtain high thermal conduction and relieve the stress from the mismatch between the coefficients of thermal expansion of YAG and copper. The copper heat sink was cooled in vacuum by LN to a temperature of 77 K.

A fiber coupled water cooled 936 nm laser diode (BWT Inc.) was used as the pump source. The core diameter of the fiber was 135 μm with a numerical aperture of 0.22. The laser diode worked in quasi-continuous wave (QCW) mode. The pump pulse duration was set to 1 ms and the repetition rate was set to 10 Hz. The pump light from the fiber was coupled to the microchip crystal with 44 μm (FWHM) diameter by a pair of lenses.

The laser output energy was measured with an energy meter (QE8SP-B-MT, Gentec-EO). The emission spectra of the laser were measured with a spectrometer (HR4000, Ocean Optics). The laser pulse characteristics were detected with a fast photodiode (<25 ps rise time, ET-3500, Electro-optics Technology) and recorded with a 12 GHz bandwidth oscilloscope (Infiniium DSO81204B, Agilent). The beam profile was measured using a charge coupled device (DMK51BU02.WG, The Imaging Source, GmbH).

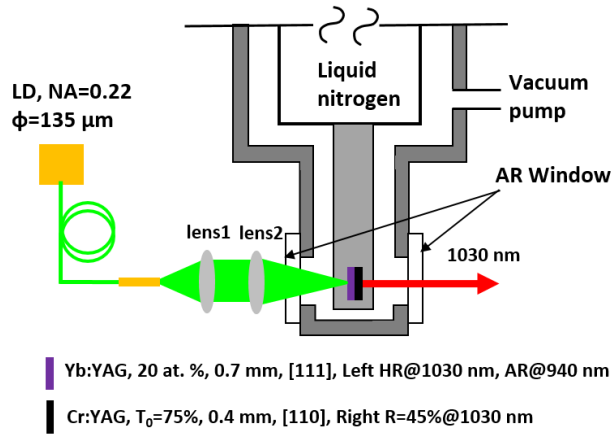

Figure 1. Experimental setup of the Yb:YAG/Cr:YAG microchip laser.

In the experiment, we found that the pulse duration is dependent on the pump focus spot position on the crystal, as shown in Figure 2 (a). The focus position of zero corresponds to the position that results in the lowest threshold pump power. Under this condition, the output laser pulse duration was as long as 182 ps. Then, we moved the pump focus spot away from the saturate absorber and the pulse duration reduced to 172 ps. There was an opposite effect if we moved the focus spot in the opposite direction. This effect was due to the pump induced bleaching of the saturable absorber. Since the microchip laser crystal is thin, the transmitted pump light from Yb:YAG is strong and will partially bleach the saturable absorber, which reduces the saturable loss for the formation of the PQS pulse and increases the minimum pulse duration. Moving the focus spot away from the saturable absorber, increases the pump intensity on the gain medium and decreases it on the saturable absorber so the pump induced bleaching decreases. As a result, the output pulse duration decreases. Zayhowski et al. [6] have made a detailed analysis of this effect both theoretically and experimentally.

To further reduce the pump induced bleaching effect and get a shorter pulse duration, we also tilt the microchip crystal. As shown in the insert in Figure 2 (b), the pump light is incident on the crystal surface at an angle (~5.2 º) to the normal. In this layout, the pump induced bleaching area in the saturable absorber is angled away from the generated laser beam's direction. Thus, we weaken the effects on the formation of the PQS pulse from pump induced bleaching. By tilting the microchip crystal, we got a pulse duration as short as 160.6 ps (FWHM). It was very close to our calculated 147.2 ps from rate equations. Figure 2 (b) shows the measured pulse profile.

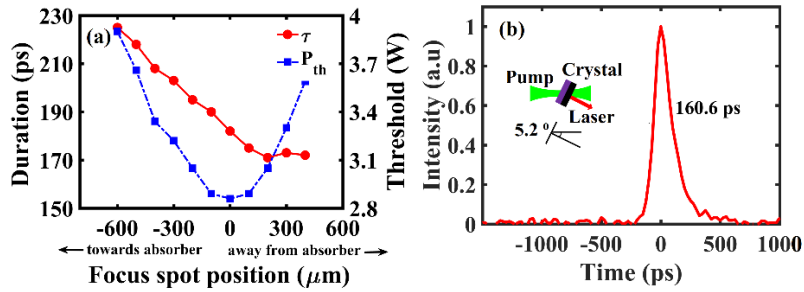

Figure 2 (a) Laser pulse duration (red) and threshold pump power (blue) as a function of pump focus spot position. Zero corresponds to the minimum threshold pump power position. (b) Measured pulse profile. The inserted figure shows the tilting of the microchip crystal.

The output pulse energy is 6.3 μJ. With a 160.6 ps pulse duration, the resulting peak power is as high as 39 kW. It is easy to enlarge the pump beam diameter and the output pulse energy can increase to over 100 μJ. However, in the experiment, with a larger pump beam diameter, the output pulse duration increased even though we tried to move the pump focus spot away from the saturable absorber. The reason is that a larger pump beam diameter will result in a longer pump beam Rayleigh length, so the pump beam diameters on the gain medium and saturable absorber are almost the same. As a result, we can't sufficiently reduce the pump induced bleaching. We think this problem can be solved by tuning the pump wavelength to exactly match the Yb:YAG peak absorption wavelength and increasing the Yb:YAG doping concentration. Figure 3 (a) shows the stability of the experimental output energy over a 10 min period. The root of mean square (RMS) variability is as small as 0.9%. This high stability can be attributed to the simple microchip structure.

Figure 3 (b) shows the measured beam radius as a function of beam propagation position. The fitted beam propagation factor $M^2$ is 1.02 and 1.01 in the horizontal and vertical directions, respectively. The inset figure is the laser focus spot. The output beam quality is quite high, which is beneficial for applications such as nonlinear frequency conversion.

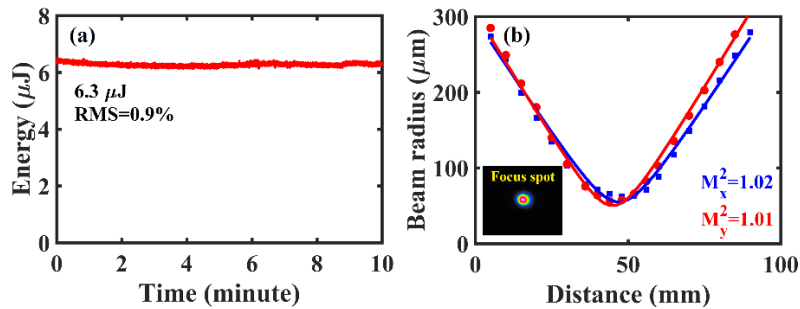

Figure 3 (a) Variability of output energy over a 10 min period. (b) Measured beam radius as a function of beam propagation position. The inserted figure is the laser focus spot.

We measured the laser output spectrum, as shown in Figure 4 (a). The center wavelength was 1029.4 nm with 0.13 nm bandwidth (FWHM). The bandwidth is near the limit of the spectrometer's resolution. The narrow bandwidth is one of the advantages of the microchip laser.

Since linear polarization is important for many applications, we measured the laser polarization using a polarizer. Figure 4 (b) shows the experimental (solid circles) and calculated (solid line) output laser intensities after the polarizer for different polarizer angles. We can see that the laser polarization is purely linear in the vertical direction.

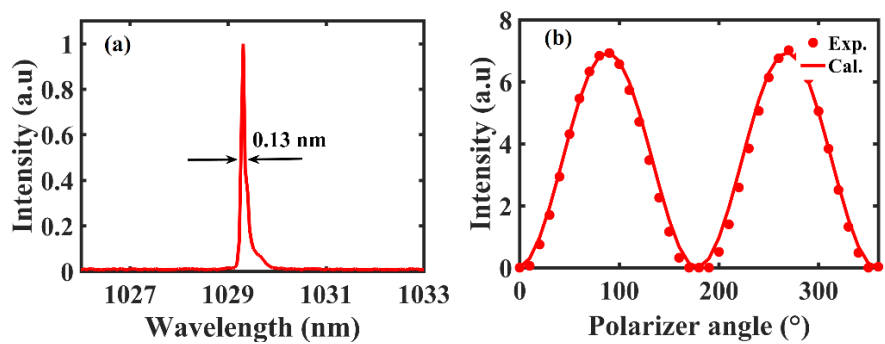

Figure 4 (a) Measured output spectrum (b) Measured and calculated laser intensities after a polarizer at different angles.

In conclusion, we have developed a diode pumped cryogenically cooled Yb:YAG/Cr:YAG microchip laser. As short as 160.6 ps pulse duration was obtained. Cryogenically cooled Yb:YAG/Cr:YAG microchip lasers are, therefore, a promising method to generate high energy ultrashort picosecond laser pulses.

Funding Information

Japan Society for the Promotion of Science (JSPS) KAKENHI (JP26287145); Photon Frontier Network of the Ministry of Education, Culture, Sports, Science and Technology (MEXT), Japan.

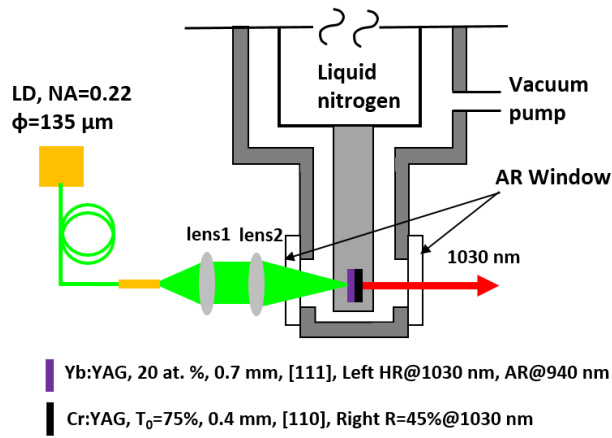

Figure 1. Experimental setup of the Yb:YAG/Cr:YAG microchip laser.

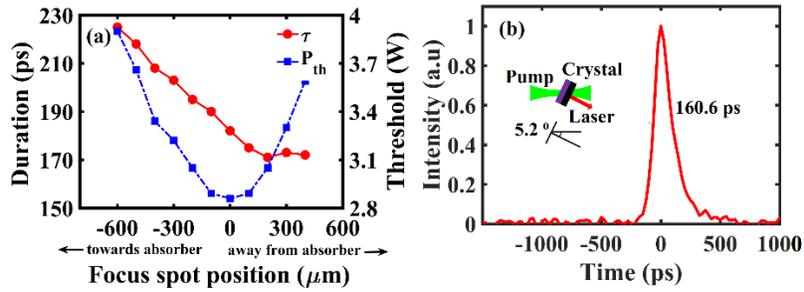

Figure 2 (a) Laser pulse duration (red) and threshold pump power (blue) as a function of pump focus spot position. Zero corresponds to the minimum threshold pump power position. (b) Measured pulse profile. The inserted figure shows the tilting of the microchip crystal.

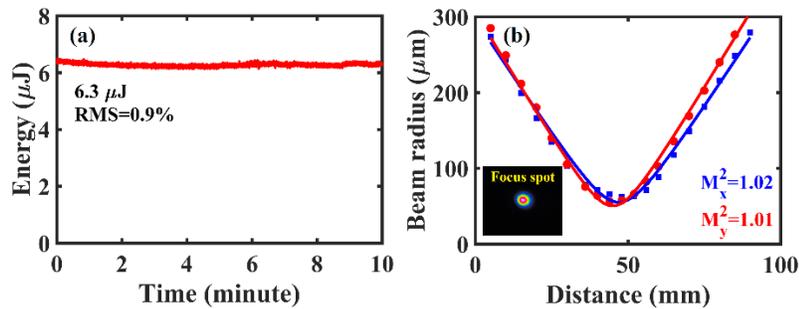

Figure 3 (a) Variability of output energy over a 10 min period. (b) Measured beam radius as a function of beam propagation position. The inserted figure is the laser focus spot.

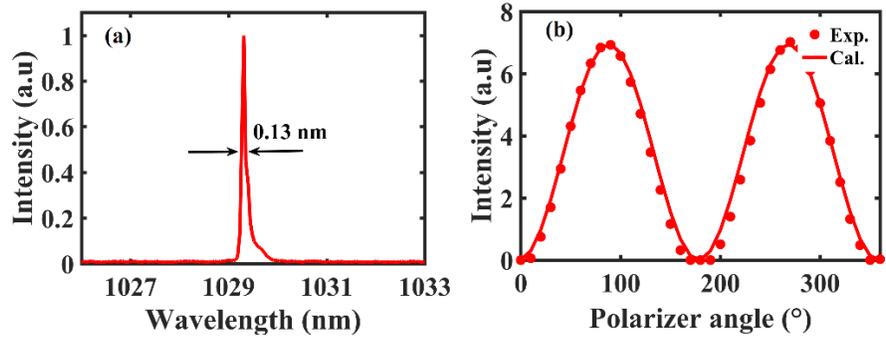

Figure 4 (a) Measured output spectrum (b) Measured and calculated laser intensities after a polarizer at different angles.